\documentstyle[aps,pre,preprint,amssymb]{revtex}
\begin{document}
\draft

\author{C. L. Emmott and A. J. Bray}

\address{Department of Physics and Astronomy, University of Manchester, 
Manchester, M13 9PL, UK.}

\title{Coarsening Dynamics of a One-Dimensional Driven Cahn-Hilliard System}

\date{\today}
\maketitle

\begin{abstract}

We study the one-dimensional Cahn-Hilliard equation with an additional 
driving term representing, say, the effect of gravity. We find that 
the driving field $E$ has an asymmetric effect on the solution for 
a single stationary domain wall (or `kink'), the direction of the 
field determining whether the analytic solutions found by Leung [J. Stat.\ 
Phys.\ {\bf 61}, 345 (1990)] are unique. The dynamics of a kink-antikink 
pair (`bubble') is then studied. The behaviour of a bubble is dependent 
on the relative sizes of a characteristic length scale $E^{-1}$, 
where $E$ is the driving field, and the separation, $L$, of the 
interfaces. For $EL \gg 1$ the velocities of the interfaces are negligible, 
while in the opposite limit a travelling-wave solution is found with a 
velocity $v \propto E/L$.  For this latter case ($EL \ll 1$) a set of 
reduced equations, describing the evolution of the domain lengths, is 
obtained for a system  with a large number of interfaces, and implies 
a characteristic length scale growing as $(Et)^{1/2}$. Numerical 
results for the domain-size distribution and structure factor confirm 
this behaviour, and show that the system exhibits dynamical scaling 
from very early times. 

\end{abstract}



\newpage

\section{Introduction}
\label{intro}
When a system is quenched from a homogeneous high-temperature phase 
into a two-phase region, domains of the new equilibrium phases form and 
evolve with time.  The dynamics of systems described by a conserved order 
parameter, such as binary alloys or binary liquids undergoing phase 
separation, are conventionally modelled by the Cahn-Hilliard equation 
\cite{Bray}. At late times after the quench, the domain coarsening is 
well described by a scaling phenomenology, with a single characteristic 
length scale $L(t)$. 

Recently there has been much interest in the dynamics of phase separation 
in the presence of an external driving field\cite{Yeung92,PPD,Alex}, 
as this has applications to, for example, spinodal decomposition in a 
gravitational field.   

In order to capture the dynamics of a system in the presence of an 
external driving field, an order-parameter-dependent diffusion 
coefficient (or `mobility') is required \cite{Yeung92,Yeung93,Kitahara}. 
The resulting modification of the Cahn-Hilliard equation has been 
studied by several authors both analytically and numerically 
\cite{Yeung92,Yeung93,Kitahara,Leung88,Leung90,Lacasta93}.
In all the two-dimensional simulations the two phases form structures which 
align along the direction of the field, with different time-dependences for 
the typical length scales parallel and perpendicular to the field. 
Analytic work on interfacial dynamics has revealed an instability in 
surfaces perpendicular to the field\cite{Yeung92,Machado} 
which explain this configuration. However, given the complexity of the 
problem, no satisfactory scaling results have been obtained.

In order to obtain some qualitative understanding, we study in this paper 
the Cahn-Hilliard equation in one dimension. 
This system has been studied in the absence of a field by several authors 
\cite{Kawakatsu,Kawasaki,Majumdar}. In the presence of a field, 
analytic solutions have been obtained for a single interface in an infinite
system \cite{Leung90}. In section \ref{sec:single} we show, however, that 
these solutions are not unique. The direction of the field relative to the 
kink profile determines whether the analytical solution obtained by 
Leung \cite{Leung90} is unique, or just one of a family of solutions.  
The work in this section motivates the introduction of an additional 
characteristic length scale inversely proportional to the field 
strength\cite{Leung88}. 

In section \ref{sec:bubble} the stationary profile of kink-antikink pairs is
studied numerically.  The behaviour of the of these systems is dependent on 
the relative values of the characteristic length ${E^{-1}}$ and $L$, the
distance between the interfaces. These systems are considered in two separate
limits, $EL \ll 1$ and $EL \gg 1$.  For $EL \ll 1$, the kink-antikink pair 
form a travelling wave, moving at a speed proportional to $E/L$. 
For $EL \gg 1$, the domain walls are essentially frozen. 

The limit $EL \ll 1$ is reconsidered in section \ref{sec:model}, in the 
context of a system with many interfaces. We consider the case in which 
all the domain lengths, $L_i$, satisfy $EL_i \ll 1$, and show that 
a simple effective dynamics for the interfaces can be constructed, in which 
the velocity of a given interface is proportional to the difference of the 
inverse lengths of the domains on either side of it. This then leads 
to a simple equation for the time evolution of the domain lengths 
themselves, from which it follows immediately that any characteristic 
length scale must grow as $\sqrt{t}$. Numerical simulations of the 
effective dynamics demonstrate scaling, with an average domain size 
growing as $\sqrt{t}$ as expected.

Section \ref{conclusions} concludes with a discussion and summary of the 
results. 
 
\section{Single-Interface Solutions of the Cahn-Hilliard equation}       
\label{sec:single}
The phase-separation dynamics of a conserved system are modelled by the 
Cahn-Hilliard equation\cite{Bray}
\begin{equation}
{\partial\phi\over\partial t} = \underline\nabla .\left (
\lambda\underline\nabla {\delta F\over\delta\phi} \right ),
\label{TDGLbare}
\end{equation}
where $F[\phi]$ is the usual $\phi^4$ Hamiltonian, but with an additional 
term to include the effects of the external field: 

\begin{equation}
F[\phi]=\int\,d^dx\left ( {1\over 2}(\underline\nabla\phi)^2
+{1\over 4}(1-\phi^2)^2 - \underline{\tilde{E}}.\underline x\,\phi\right ).
\label{FGL}
\end{equation} 

Throughout this paper only the deterministic Cahn-Hilliard equation is
considered, i.e., the Langevin noise term has been omitted.  This 
restricts the limits of validity of these results to temperature 
$T=0$ or, more generally, to $T$ small enough that the effects of 
thermal fluctuations may be neglected.

If the mobility $\lambda$ is independent of the order parameter, inserting 
(\ref{FGL}) into ({\ref{TDGLbare}) shows immediately that the term involving 
$\tilde{E}$ drops out of the dynamics. For non-trivial (and physical) 
results, therefore, the inclusion of an external field requires that the 
dependence of the mobility $\lambda$ on the order-parameter field $\phi$ 
should be explicitly taken into account \cite{Yeung92,Yeung93,Kitahara}. 
In this paper we take the simplest form for $\lambda$ that maintains 
the symmetry under $\phi \to -\phi$, i.e.\ $\lambda=(1-a\phi^2)$. 
Hence rewriting equation (\ref{TDGLbare}) in one dimension, and defining 
$a\tilde{E}=E$, we obtain
\begin{equation}
{\partial\phi\over\partial t} = {\partial^2\mu\over\partial x^2} + 
 E{\partial(\phi^2)\over\partial x}=-{\partial J\over\partial x},
\label{cahn}
\end{equation}
where 
$$\mu=\phi(\phi^2-1)-\phi''$$ 
is the chemical potential and $J$ is the current. 
Note that we have only kept the leading-order term in $a$, as is 
conventional. Technically, this is equivalent to taking the 
limits $a \to 0$, $\tilde{E} \to \infty$, holding $E = a\tilde{E}$ fixed.

For an infinite system, analytic solutions to equation (\ref{cahn}) have  
been found for a single stationary interface \cite{Leung90}.  
The order-parameter profile for a kink solution is given by
\begin{equation}
\phi (x) = \phi_\infty\,\tanh (\phi_\infty x/\sqrt{2}),
\label{TDGL_1D}
\end{equation}
with 
\begin{equation}
\phi_\infty^2 = 1 + \sqrt{2}E,
\label{exact} 
\end{equation}
valid for both signs of $E$. The corresponding antikink solution is 
obtained by inserting an overall minus sign in (\ref{TDGL_1D}), and 
replacing $E$ by $-E$ in (\ref{exact}). [This reflects the symmetry 
of (\ref{cahn}) under $\phi \to -\phi$, $E \to -E$.]

In general, solutions to non-linear equations are not unique. 
The aim of this section is to investigate the solutions of equation
(\ref{cahn}) numerically, and to present some simple analysis which 
accounts for the (at first sight surprising) results.
                                          
The continuum equation (\ref{cahn}) was  discretized on the interval 
$(-L/2,L/2)$, and antiperiodic boundary conditions, $\phi(-L/2) = 
-\phi(L/2)$, were imposed. A kink solution was sought by using an 
antisymmetric tanh profile, with $\phi(0)=0$ and $\phi (L/2) > 0$, as 
the initial condition. A range of initial values of $\phi(L/2)$ was used. 
This initial state was then evolved under the dynamics (\ref{cahn}) until 
the system reached a stationary profile. A system size $L=500$ was used 
with a mesh size of $0.5$. This value of $L$ was found to be large enough 
that the results showed no finite-size effects. The time step used in the 
iteration was $\Delta t = 0.005$. 

It was found that the number and form of the stationary profiles found was 
strongly dependent on the sign of the field. For each negative field value 
a unique stationary wall profile was found, independent of the initial 
conditions used, i.e.\ of the initial value of $\phi(L/2)$.  The final 
value of $\phi(L/2)$ obtained was found to be below the analytically 
predicted value $(1+\sqrt{2}E)^{1\over 2}$ [equation (\ref{exact})]; 
the discrepancy vanished, however, as the mesh size was decreased (see 
figure 1).                                   

For a positive field a family of solutions was found, with the final 
value of $\phi(L/2)$ equal to the initial value. 
The interface was found in general to be broader than for $E<0$, the 
asymptotic exponential decay to the value $\phi(L/2)$ occurring at a rate 
determined by the magnitude of the field, the decay rate decreasing as 
the field decreases. For larger values of the field, this exponential 
tail became oscillatory (see figure 2).                                   

Some simple analysis of the differential equation (\ref{cahn}) can be 
used to explain some of these features. The existence of a solution 
representing a single interface in an infinite system is dependent on 
the choice of boundary conditions. For a physical solution, two of the 
boundary conditions imposed must be $\phi(0)=0$, which 
fixes the position of the interface,  and $\phi''(0)=0$, a consequence of
the anti-periodicity.  The other two boundary conditions, $\phi'(0)$ and 
$\phi(\infty )$, must be chosen to ensure the existence of a solution.  
This places constraints on the the possible values of these
variables. By the use of some simple stability analysis it is possible to 
deduce the type of constraints this places on $\phi'(0)$ and $\phi(\infty )$
and hence the number of values for which a solution exists.

Equation (\ref{cahn}) can be integrated once to give a 
third order differential equation, 
\begin{equation}
E\phi ^2 + (3\phi ^2-1)\phi' - \phi''' = E\phi_\infty^2\ , 
\label{third}
\end{equation} 
the integration constant $E\phi_\infty^2 \equiv E\phi^2(\pm \infty)$ 
ensuring that $\phi(x)$ tends to a constant value at infinity. 
Linearizing around this value for large positive $x$, 
$\phi(x) = \phi_\infty +\tilde{\phi}(x)$, gives        
\begin{equation}
\tilde{\phi}''' + (1 - 3\phi_\infty^2)\tilde{\phi}' 
- 2E\phi_\infty \tilde{\phi} =0.
\end{equation}
It follows that $\tilde{\phi}$ will have solutions of the form 
$\exp(\lambda x)$, where the possible values of $\lambda$ are the  
roots of the cubic equation
\begin{equation}
\lambda^3 + (1 - 3\phi_\infty^2)\lambda = 2E\phi_\infty\ .
\label{cubic}
\end{equation}
The nature of the roots of this equation is dependent on the
values of $\phi_\infty$ and $E$.  

Note that in the present discussion we are taking $\phi(\infty)>0$, 
and considering both signs of $E$. This means that we are dealing with 
a kink, not an antikink. Later, when we consider systems with both 
kinks and antikinks, we will fix $E>0$. Since, however, the equation of 
motion (\ref{cahn}) is invariant under $E \to -E$, $\phi \to -\phi$, 
it follows that the properties of an antikink for $E>0$ are identical 
to those of a kink for $E<0$. In the following stability analysis, 
therefore, we restrict the discussion to kinks. 

There are four cases to consider, as illustrated in figure 3.  
\renewcommand{\theenumi}{\alph{enumi}}
\renewcommand{\labelenumi}{\theenumi)}
\begin{enumerate}     

\item  $E>0$ and $(E\phi_\infty)^2 > (\phi_\infty^2-{1\over 3})^3$ 

Two roots are complex with negative real parts, and the third root 
is real and positive. 

\item  $E>0$ and $(E\phi_\infty)^2 < (\phi_\infty^2-{1\over 3})^3$

All roots are real -- one positive and two negative. 
For $E \to 0$, $\phi_\infty \to 1$ and the roots are 
$\lambda \approx \pm \sqrt{2}$, $-E$.

\item  $E<0$ and $(E\phi_\infty)^2 < (\phi_\infty^2-{1\over 3})^3$

All roots are real -- one negative and two positive. 
For $E \to 0$, the roots are  
$\lambda \approx \pm \sqrt{2}$, $-E$.

\item  $E<0$ and $(E\phi_\infty)^2 > (\phi_\infty^2-{1\over 3})^3$

Two roots are complex with positive real parts, and the third root is 
real and negative. 

\end{enumerate}
               
For $x \to \infty$, the function $\tilde{\phi}(x)$  has the form 
$$ \tilde{\phi}(x) = \sum_{i=1}^3 a_i\,\exp(\lambda_i x). $$
The coefficients $a_i$ are fixed by the initial conditions and the 
value of $\phi_\infty$. Linear stability of the putative solution
with $\phi(x) \to \phi_\infty$ as $x \to \infty$ requires that 
$\tilde{\phi}$ vanish at infinity, i.e.\ the coefficients 
$a_i$ corresponding to eigenvalues $\lambda_i$ with positive real part 
must vanish. 

To see what this implies, consider integrating the third-order 
equation (\ref{third}) with the initial conditions $\phi(0)=0=\phi''(0)$, 
which follow from the antisymmetry of the solution, and some fixed 
$\phi'(0)$. One also needs to specify the value of the parameter 
$\phi_\infty$ appearing in the equation. The two parameters $\phi'(0)$ 
and $\phi_\infty$ must be varied until the solution $\phi(x)$ 
approaches the same value $\phi_\infty$ as $x \to \infty$. If the 
input values deviate slightly from the correct values, the solution 
will ultimately diverge from $\phi_\infty$ due to the presence of 
unstable modes (with $Re\,\lambda_i > 0$) in the linearised 
solution for $\tilde{\phi}$. To ensure that such diverging terms are 
absent, the parameters $\phi'(0)$ and $\phi_\infty$ have to be adjusted 
until the corresponding coefficients $a_i$ vanish. This is the familiar 
`shooting method' of solving non-linear differential equations. 

For $E>0$ there is only one unstable mode.  Hence for solutions to 
exist only one coefficient must vanish, imposing a constraint of the form 
$a_1(\phi'(0),\phi_\infty)=0$.  This defines a line in parameter space 
for which this coefficient vanishes, resulting in a range of possible 
values for $\phi_\infty$. One therefore expects a family of solutions, 
parametrised by $\phi_\infty$, for $E>0$. 

For $E<0$, however, there are two roots with positive real parts. Therefore 
the coefficients of both unstable modes are required to vanish.  Hence the 
only choice for the parameters $\phi_\infty$ and $\phi'(0)$ for which a 
solution exists is at the intersection of the two lines defined by setting 
both coefficients to zero, i.e.\ 
$a_1(\phi'(0),\phi_\infty)=0=a_2(\phi'(0),\phi_\infty)$.  
This results in a unique solution for $E<0$. 

The values of $E$ and $\phi_\infty$ determine whether the roots of the 
cubic equation (\ref{cubic}) are real or complex.  The existence of 
complex roots with negative real parts for $E$ sufficiently large 
and positive [case (a) in figure 3] results in oscillatory 
behaviour.

For a given stable solution, the eigenvalue $\lambda$ with 
the least-negative real part determines the rate of the asymptotic 
exponential decay towards $\phi_\infty$. For $E$ small and positive, 
this eigenvalue is $-E$. Hence for $E>0$, the value of $E$  
determines a characteristic length in the system, $E^{-1}$, 
which defines the width of any boundary effect at an interface. This
characteristic length is important when considering systems with more than
one interface as in the following sections. For $E<0$, by contrast, 
there is only one negative eigenvalue, which approaches a non-zero 
limiting value for $E \to 0$: the width of a kink for $E<0$ 
therefore remains of order unity for small $E$. 

How do we interpret the analytic kink solution (\ref{TDGL_1D}) in the light 
of the foregoing discussion? For $E<0$, the analytic solution is the 
unique solution we predict. For $E>0$, we have argued for (and numerically 
demonstrated) a family of solutions. Which one corresponds to equation 
(\ref{TDGL_1D})? It is easy to show that the exponential decay of 
(\ref{TDGL_1D}) to $\phi_\infty$ as $x \to \infty$ is governed by the 
more negative of the two negative eigenvalues arising from the stability 
analysis. The exact solution (\ref{TDGL_1D}) therefore corresponds to 
the case where the amplitude $a$ corresponding to the least negative 
eigenvalue vanishes, i.e.\ to just one member of the family of solutions. 

Throughout this section we have considered only kink solutions 
($\phi>0$ at positive infinity) and allowed $E$ to take either sign. 
We conclude by interpreting our results for the physically relevant case 
where the sign of $E$ is fixed (we take $E>0$) and both kinks and 
antikinks are present. From the symmetry of the dynamics under 
$\phi \to -\phi$, $E \to -E$, it follows that for $E>0$ there is a 
family of kink solutions, but a unique antikink solution. In a system 
with well-separated (relative to $E^{-1}$)  kinks and antikinks, however, 
continuity of the function $\phi$ will select the kink solution that 
matches smoothly on to the unique antikink solution. When the 
kink-antikink the separation is small compared to $E^{-1}$, more 
interesting behaviour is possible. This is the subject of the next section. 

\section{Single-Bubble Solutions of the Cahn-Hilliard equation}       
\label{sec:bubble}                  
In this section solutions to the Cahn-Hilliard equation for a kink-antikink 
pair in a periodic system are considered. The kink-antikink pair 
is a bubble of plus phase moving in a sea of minus phase. 
The velocities of the interfaces were investigated for various field 
strengths and kink-antikink separations.

The work in the previous section produced a characteristic
length ${E^{-1}}$, which determines (for a kink) the width of any boundary 
effect present at an interface.  In the systems considered in this section 
there is a second important characteristic length, $L$, the distance between 
the two interfaces.  The behaviour of these systems is dependent on 
the relative values of these two characteristic lengths. In this section 
we will consider two limits, firstly the case where $EL \gg 1$ and 
then the opposite limit, $EL \ll 1$.
The results from the second regime motivate a simple model for the
many interface problem which is dealt with in section {\ref{sec:model}}.    
                  
\subsection{$EL \gg 1$}
The numerical solutions found for a kink-antikink pair in this limit are
stationary.  The value of the order-parameter between interfaces was 
dependent only on the value of the field and was numerically found to be 
consistent with $\phi\approx\pm(1-\sqrt{2}E)^{1\over 2}$, the value given 
for an antikink by the analytical solution (\ref{TDGL_1D}). Some examples 
of the profiles found are shown in figure 4.  

These results can be explained using the results of section \ref{sec:single}. 
If we assume that the system will reach a steady state and the
order-parameter may be written as a travelling wave, $\phi=\phi(x-vt)$,
then the Cahn-Hilliard equation (\ref{cahn}) may be integrated once to obtain
\begin{equation}
E\phi^2 + (3\phi^2-1)\phi' - \phi''' = J_o -v\phi,
\label{TDGL_trav}
\end{equation} 
where $J_o$ is a constant of integration and $v$ is the wave velocity.
In the limit $EL \gg 1$ the order-parameter is approximately constant 
both inside and outside the bubble, $\phi\approx\pm(1-\sqrt{2}E)^{1\over 2}$.  
Inserting this into the above equation in these regions,  and demanding 
that it hold inside domains of both sign, it is clear that one must 
have $J_0 = E(1-\sqrt{2}E)$ and $v=0$. 

Therefore the solution is stationary, and since in this limit the separation 
$L$ of the interfaces is large compared to $E^{-1}$, the profile is the 
combination of two single interface solutions. 

For one of the interfaces there exists a family of solutions for a fixed value
of the field whereas the other interface has a unique solution with a plateau 
height given by $\pm(1-\sqrt{2}E)^{1\over 2}$. This specifies the
boundary conditions for the first interface, hence picking out one from the 
family of solutions. 

\subsection{$EL \ll 1$}   
In this limit the system was found numerically to evolve to a steady state 
with a fixed profile moving at a finite velocity.  This may be shown by a 
comparison of the order-parameter with the current, since for a travelling 
wave solution equation (\ref{cahn}) gives $-v\phi'=-J'$.  Therefore the 
order-parameter will be a multiple of the current $J$, plus an overall 
constant. This expectation is confirmed by the data presented in figure 5.

As in the previous limit, a long-range boundary effect exists at the kink 
interface, with the order-parameter decaying exponentially to a fixed value 
outside the bubble. However, since $EL \ll 1$ the full exponential 
tail on the the other side of the interface (i.e.\ between the two interfaces) 
is missing, and the order-parameter varies roughly linearly between the two 
interfaces, through a height difference proportional to the field. 
This linear regime between the interfaces is evident in figure 5.

Using the assumption (verified numerically) that the solution may be written 
as a travelling wave,  equation (\ref{TDGL_trav})
becomes
\begin{equation}
E\left (\phi^2-\phi^2_\infty \right ) + (3\phi ^2-1)\phi' - \phi''' +
v\left (\phi+\phi_\infty \right ) = 0.
\label{trav}
\end{equation} 
where $-\phi_\infty$ is the value of the order-parameter in the minus 
phase far from the interfaces. The integration constant $J_0$ in 
(\ref{TDGL_trav}) was fixed in (\ref{trav}) by the requirement 
$\phi(\pm \infty) = -\phi_\infty$. 
Integrating this equation over the whole system yields an expression for the
velocity: 
\begin{equation}
v= E\,{\int\,dx\,(\phi^2_\infty-\phi^2) \over
		 \int\,dx\,(\phi_\infty+\phi)} 
 = {E\over 2\phi_\infty L}\,\int\,dx\,(\phi^2_\infty-\phi^2)\ ,
\label{vel}
\end{equation}                 
where the final result follows from noting that $\phi \simeq \phi_\infty$ 
inside the bubble and $-\phi_\infty$ outside. 

The function $(\phi^2_\infty-\phi^2)$ is peaked at the interfaces. 
This can be used to explain why a finite velocity is expected.
Consider first the case $EL \gg 1$, for which we showed above that 
the velocity is zero. For this case, therefore, 
$\int\,dx\,(\phi^2_\infty-\phi^2)$ must be zero. Since the kink and 
antikink are well separated, and the order parameter saturates at the 
value $\phi_\infty$ in the intermediate plateau region, 
$\int\,dx\,(\phi^2_\infty-\phi^2)$ is a sum of kink and antikink 
contributions. One of the interfaces, corresponding to the antikink, is known 
analytically [from equation (\ref{TDGL_1D})], and the contribution it makes 
to the integral is $2\sqrt{2}\phi_\infty$. The contribution from the kink 
interface, therefore, must give an overall negative contribution of 
$-2\sqrt{2}\phi_\infty$. This is possible because, as is clear from 
figure 4, the kink solution overshoots $\phi_\infty$ and 
approaches the plateau at $\phi_\infty$ from the `wrong' side. The function 
$\phi_\infty^2 - \phi^2$ for the kink therefore contains both positive 
and negative regions, as shown in figure 6. 
The negative regions evidently contain the 
larger area [note that, although the magnitude of $\phi_\infty - \phi^2$ 
in these regions is $O(E)$, the decay constant is also $O(E)$, leading 
to an area of $O(1)$], in such a way that the integral for the kink 
interface is precisely $-2\sqrt{2}\phi_\infty$. 

Consider now the case $EL \ll 1$. In this case, the interior of the 
bubble is too small to contain the full exponential tail from the kink.  
The contributions from the kink to $\int (\phi_\infty^2 - \phi^2)$ will 
therefore be less negative than when $EL \gg 1$, resulting in a net positive 
value for the integral and, via (\ref{vel}), a positive bubble velocity,  
$v \propto E/L$. Therefore in this limit the interfaces evolve to a fixed 
profile which moves with a velocity $v \propto E/L$. The qualitative 
tendency of smaller bubbles to move faster than larger ones was noted by 
Yeung et al., on the basis of numerical studies\cite{Yeung92}. 

We can make this argument quantitative in the limit of small $E$, 
for which $\phi_\infty \to 1$.  
In this limit, the area under the kink peak in figure 6 
approaches the same value, $2\sqrt{2}$, as the area under the antikink. 
The negative region between the two peaks is completely suppressed, as 
the decay length, $1/E$ is much larger than the separation, $L$, of the 
peaks. The negative contribution from the semi-infinite region 
to the left of the kink peak, however, retains the value $-2\sqrt{2}$. 
The total area under the kink is therefore zero, and the area under the 
kink-antikink pair is $2\sqrt{2}$. Equation (\ref{vel}) then gives 
a velocity $v=\sqrt{2}\,E/L$ for $E \to 0$.  Applying this result to 
the data of figure 5, for which $E=0.01$ and $L=20.88$ (measured 
between the zeros of $\phi$) gives $v=6.77 \times 10^{-4}$, which compares 
well with the measured velocity of $6.82 \times 10^{-4}$. 

The results of this section will be used in the following section to 
derive the set of reduced equations employed to simulate a multi-domain 
system. As a first step in generalising this approach to many domains, we 
express the single-bubble calculation in the limit $EL \ll 1$ in terms 
of currents. From equation (\ref{cahn}), the current $J$ can be written, 
up to an overall constant, as 
\begin{equation}
J = -\mu' - E\phi^2 = \phi''' + (1-3\phi^2)\phi' - E\phi^2. 
\label{current}
\end{equation}
The velocity of an interface is related to the discontinuity in the current 
at the interface\cite{Bray}, $v=\Delta J/\Delta \phi$, where $\Delta J$ 
and $\Delta \phi$ are the discontinuities in the current and the order 
parameter respectively. For a bubble that moves without change of shape 
(a travelling wave), $J$ must have different values $J_{in}$ 
and $J_{out}$ inside and outside the bubble. To lowest order in $E$, 
we can write $\phi^2 = \phi_\infty^2 = 1$ both inside and outside, so that 
(\ref{current}) becomes $J = \phi''' - 2\phi' - E$ in both regions. 
Outside the bubble, far from the interfaces, $\phi$ is constant, so 
$J_{out} = -E$. Inside, the order parameter changes by an amount of order 
$E$ over a distance $L$, so we estimate $\phi'= -E/L$ (and $\phi'''$ is 
negligible), giving $J_{in} \sim -E + O(E/L)$ correct to $O(E)$. 
Finally, both interfaces have the same velocity $v \sim E/L$, in 
agreement with our earlier result (\ref{vel}) based on the travelling 
wave solution. This second approach is, however, readily generalisable 
to the multi-domain situation in the small-$E$ limit.  

\section{Many Interface Model}
\label{sec:model}
In this section a set of reduced equations is derived using the work of the 
previous section to the describe the dynamics of a large number of interfaces 
in a periodic system. This reduced dynamics is numerically simulated, and the 
average domain length, domain-size distribution function and structure factor 
are calculated.  

We consider the case in which $EL_i \ll 1$ for all domains $i$, where 
the $L_i$ are the domain lengths. From the arguments of the previous section, 
we expect that, in this limit, $\phi' \sim - E/L_i$ in domain $i$, and that 
the current $J_i$ in domain $i$ is given, up to an additive 
constant, by $J_i \propto E/L_i$. The proportionality constant 
in this relation is the same for all domains (an extension of the argument 
given above for the kink-antikink pair \cite{Note} gives its value as 
$4\sqrt{2}$). 

Let us define interface $i$ to have domain length $L_i$ to its right, and 
$L_{i-1}$ to its left. If the interface is a kink ($\Delta \phi >0$) its 
velocity $v_i$ is 
\begin{equation}
v_i^k = \frac{\Delta J}{\Delta \phi} \propto E\left(\frac{1}{L_i} 
- \frac{1}{L_{i-1}}\right) .
\end{equation}
For an antikink ($\Delta \phi < 0$) the velocity $v_i^{ak}$ is given by 
the same expression, but with a factor (-1). If we arbitrarily assign 
even $i$ to domains of positive $\phi$, the equation for the time 
evolution of the domain length, $dL_i/dt = (v_{i+1} - v_i)$, becomes   
\begin{equation}
{dL_i\over dt} =
  (-1)^i\,E\,\left({1\over L_{i-1}} - {1\over L_{i+1}} \right), 
\label{sim}
\end{equation}
where an overall constant has been absorbed into the timescale. 
It follows immediately that, if $L$ is a typical domain length, 
$dL/dt \sim E/L$, i.e.\ $L(t) \sim (Et)^{1/2}$. 

The dynamics of the system were studied numerically using this set of 
reduced equations. The system was initially prepared with $100,000$ domains 
with approximately equal amounts of plus and minus phase. The field $E$ was 
absorbed into the time scale, i.e.\ we set $E=1$ in (\ref{sim}). 
The time step used in the iteration was $\Delta t \propto t$.  This was 
chosen because we expect $t^{1\over 2}$ (i.e.\ power-law) growth. Then 
the $\Delta t \propto t$ leads to $\Delta L \propto L$, i.e.\ domains 
typically change length by a fixed (small) fraction of the mean length in 
each update. Similarly, domains were annihilated if the distance between 
interfaces became smaller than a specified (small) fraction of the average 
domain size. In the calculation of the average domain size and the 
domain-size distribution function, results were averaged over $10$ runs.  
It was found that the average domain size, $\langle L(t) \rangle$, grew 
as $\sqrt{t}$ as expected, (figure 7). 
The simulation results show that the distribution function $P(L,t)$ 
(the fraction of domains which have size $L$ at time $t$) scales well 
from quite early times (figure 8). The scaling function was 
found to be linear at the origin, with a gaussian tail. However, we 
were unable to calculate it analytically. 

Figure 9 shows the structure factor. This was taken as an 
average of 500 runs.  The scaling collapse is good except near the peak, 
where the data are noisy. A ln-ln plot (figure 10) 
shows that in the limit $k\langle L \rangle \gg 1$ the structure factor 
possesses the expected Porod tail, $S(k,t)\propto k^{-2}$. The constant of 
proportionality is determined by the average domain-wall density\cite{Bray}, 
$S(k,t)=4\rho k^{-2}$.  Hence we would expect a `Porod plot' of 
$k^2 \langle L \rangle S(k,t)$ to tend to the value $4$ at large 
$k\langle L \rangle$. The data, although noisy in this regime, are 
consistent with this expectation. 

The small-$k$ data in figure 10 suggest the  
quadratic dependence, $S(k,t) \propto k^2$. We can show that 
this is the correct small-$k$ behaviour, following the methods used to 
establish the $k^4$ small-$k$ behaviour\cite{Bray,Furukawa} for the 
non-driven Cahn-Hilliard equation in dimensions $d>1$. 

For $k \to 0$, equation (\ref{cahn}) can be written in Fourier space as 
\begin{equation}
\frac{d\phi_k}{dt} = iEk (\phi^2)_k\ ,
\end{equation}
where the term of order $k^2$ is negligible and has been omitted. Integrating 
with respect to $t$, multiplying by $\phi_{-k}$, and averaging gives the 
structure factor 
\begin{equation}
S(k,t) = -S(k,0) + 2\langle \phi_k(t)\phi_{-k}(0) \rangle 
  + E^2 k^2\int_0^t dt_1 \int_0^t dt_2\,\langle (\phi^2[t_1])_k 
(\phi^2[t_2])_{-k} \rangle\ .
\label{small-k}
\end{equation}
Since $S(k,t)$ has the scaling form $S(k,t)= Lg(kL)$, where $L$ is a 
shorthand for $\langle L \rangle$, it follows that $S(k,0)$ is negligible 
compared to $S(k,t)$ at late times (large $L$) and can be neglected. 
Similarly, $\langle \phi_k(t)\phi_{-k}(0) \rangle$ has the scaling 
form\cite{Bray} $L^{\lambda}h(kL)$, where $\lambda < d/2=1/2$ follows 
from the Cauchy-Schwarz inequality\cite{YRD}, so this term can also 
be dropped for large $L$. 

Next we consider the equal-time correlation function 
$D(r,t) = \langle [\phi_\infty^2 - \phi^2(x)] 
[\phi_\infty^2 - \phi^2(x+r)] \rangle$. It's Fourier transform is the 
equal-time version of the quantity required in (\ref{small-k}). The 
function $\rho(x) \equiv \phi_\infty^2 - \phi^2(x)$ vanishes except near 
domain walls, so it is essentially a domain-wall density function, with 
delta-function contributions from the walls \cite{Note}. It follows that 
$\langle \rho \rangle \sim 1/L$, the mean domain-wall density, while 
$D(r,t)$ has the scaling form $D(r,t) = L^{-2}f_D(r/L)$. The two-time 
generalisation is $D(r,t_1,t_2) = L_1^{-2}f_D(r/L_1,L_2/L_1)$, where 
$L_1 \equiv L(t_1)$. The spatial Fourier transform required in 
(\ref{small-k}) has the form $T(k,t_1,t_2) = L_1^{-1}g_D(kL_1,L_2/L_1)$, 
which reduces to $L_1^{-1}h(L_2/L_1)$ in the limit $k \to 0$ [we do not 
expect $T$ to vanish in this limit, because $\rho$ is not a conserved 
quantity]. 

If $L(t)$ grows as a power of $t$, the double time integral in 
(\ref{small-k}) gives, up to constants, $t^2/L$, so the right-hand 
side of (\ref{small-k}) is of order $E^2 k^2 t^2/L$. Requiring that 
$S(k,t) = Lg(kL)$ have the same small-$k$ form gives 
$S(k,t) \sim k^2 L^3$, and hence $L \sim (Et)^{1/2}$. It is reassuring 
that this form for $L(t)$ agrees precisely with that obtained from the 
reduced dynamics (\ref{sim}).

\section{Conclusions}
\label{conclusions}

We have studied, analytically and numerically, the effect of an external
driving field on the coarsening dynamics of the one-dimensional 
Cahn-Hilliard equation at $T=0$. 

For a single stationary interface, it was shown that the direction of the 
field for a domain wall of a given sign determines whether there is a 
unique solution or a family of solutions. In the latter case, the approach 
of the interface profile function to its asymptotic value is governed 
by an exponential tail with a decay constant which vanishes linearly 
with the driving field $E$. There is therefore a new characteristic 
length scale, $E^{-1}$, in this system. 
                               
The behaviour of a kink-antikink pair (`bubble') falls into two classes, 
characterised by the relative values of the interface separation, $L$, and 
the new characteristic length, $E^{-1}$.  In the limit $EL \gg 1$,
the bubble profile is stationary. In the opposite limit, $EL \ll 1$, a 
bubble of plus phase moves through the minus phase with a velocity 
$v \propto E/L$.

For the many-domain coarsening dynamics an equation of motion for 
the domain lengths was derived [equation (\ref{sim})], in the which the 
length of a given domain changes at a rate determined by the lengths of 
the domains on either side. The mean domain size grows as $\sqrt{Et}$. 
Despite the apparent simplicity of this model, we have so far been 
unable to make further analytical progress. 
Numerical simulations, however, demonstrate dynamic 
scaling (for the domain-size distribution and the structure factor),  
and confirm the predicted $\sqrt{t}$ growth law. An argument for the 
observed $k^2$ behaviour of the structure factor at small $k$ has 
been presented. 

In a related work, the 1-D Ising model with Kawasaki spin-flip dynamics,  
biased in one direction, has been studied\cite{Cornell}. Numerical results 
were obtained for a range of volume fractions, including the case of 
equal volume fractions simulated here, and $\sqrt{t}$ growth demonstrated. 
There, as here, rather general arguments lead to $\sqrt{t}$ growth 
for all volume fractions -- for the present model from dimensional 
analysis of equation (\ref{sim}). For the Ising model, however, exact 
results for the domain-size distribution (and for the `persistence 
exponent'\cite{Cornell}) were obtained in the limit where one phase 
occupies a negligible volume fraction, so it is interesting to consider 
the present model in this same limit. 

When one phase (the plus phase, say) occupies a very small volume fraction, 
the system consists of domains of plus phase separated by (typically)
much larger domains of minus phase. To a first approximation, therefore, 
each plus domain can be treated as an isolated bubble in the sense of 
section \ref{sec:bubble}. Each such bubble then moves at a speed $E/L$, 
where $L$ is its length (we are taking $EL \ll 1$ here), all bubbles 
moving in the same direction. As a result, small bubbles catch up with 
larger bubbles, with which they then merge, and the combined domain 
slows down. Just before the bubbles merge, the approximation of treating 
them as independent breaks down, when the size of the intervening minus 
domain becomes comparable with the size of the bubbles. This presumably 
leads to negligible corrections to the scaling functions, however, in the 
small volume-fraction limit. This new model is so simple that analytic 
progress may be possible. However, we will leave this question for future 
work. 

\section{Acknowledgments}
CLE would like to thank Sarah Phillipson for useful discussions, and 
EPSRC (UK) for support.

\newpage
\begin{center}
FIGURES
\end{center}

{\bf FIG1}. Stationary single-interface solutions of equation (\ref{cahn}).  
There is a unique solution for each $E<0$.   
(a) $E=-0.1$, (b) $E=-0.3$, (c) $E=-0.5$. 

{\bf FIG2}. Stationary single-interface solutions of equation (\ref{cahn}).
(a) $E=0.2$, (b) and (c) $E=0.3$: two examples from a family of 
solutions. For each value of the field there is a different decay length.

{\bf FIG3}. Graphical solution (schematic) of equation (\ref{cubic}) for 
the four cases discussed in the text. 

{\bf FIG4}. Kink-antikink solutions with $L \approx 107$ and $L \approx 161$, 
for $E=0.1$, corresponding to the limit $EL \gg 1$.

{\bf FIG5}. Comparison between the order-parameter profile and the 
current for a kink-antikink pair of size $L \approx 21$, for $E=0.01$, 
corresponding to the limit $EL \ll 1$. 

{\bf FIG6}. Plot (schematic) of $\phi_\infty^2 - \phi^2(x)$ for a 
kink-antikink pair in the limit $EL \gg 1$. The total area under 
the curve is zero. 

{\bf FIG7}. Time dependence of the square of the average domain length 
for the dynamics described by equation (\ref{sim}). The mean domain 
size grows as $t^{1/2}$. 

{\bf FIG8}. Scaled domain-size distribution function for the dynamics 
described by equation (\ref{sim}).  The data correspond 
to times 10 ($\square$), 20 (+), 40 ($\ast$), 80 ($\bigcirc$), 
160 ($\times$), 320 ($\vartriangle $), 640 ($\lozenge$).

{\bf FIG9}. Scaled structure factor for the dynamics described by 
equation (\ref{sim}).  The data correspond to times between 
10 and 640, as in figure 8. 

{\bf FIG10}. Double-logarithmic plot of the scaled structure factor. 
Times between 10 and 640 are shown, as in figure 8. 
The straight lines in the small-$k\langle L \rangle$ and 
large-$k\langle L \rangle$ regimes have gradients $2$ and $-2$ 
respectively. The latter is the expected Porod regime; an argument for 
the former is given in the text. 


\begin{references}
\bibitem{Note} For the many-interface situation considered here, the 
exponential tails of the kink profiles are cut off on both sides by the 
neighbouring antikinks (for $EL \ll 1$), so the function $\phi_\infty^2 
-\phi^2(x)$ has peaks with positive (and essentially the same) area for 
both kinks and antikinks.
\end{references}
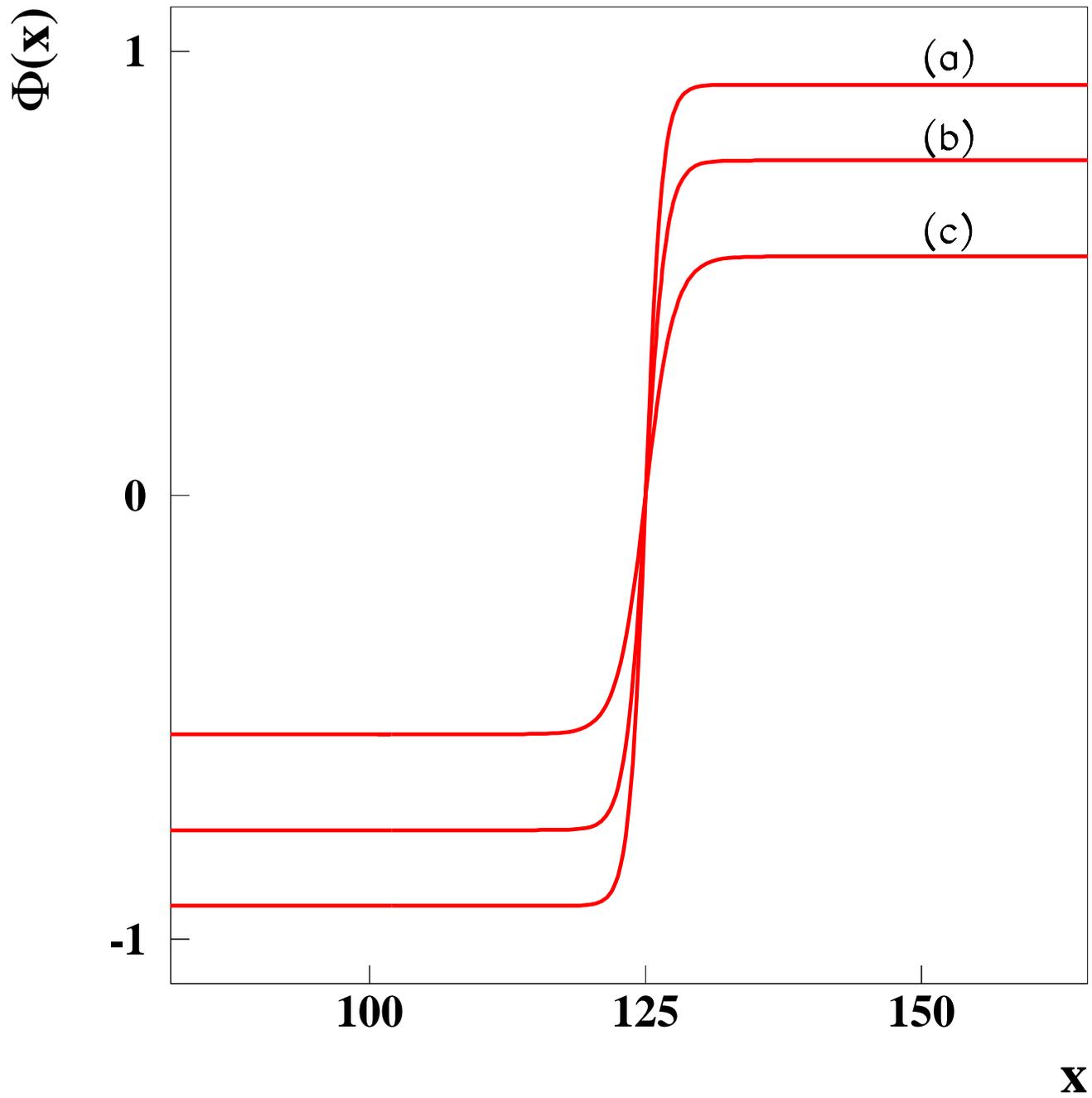
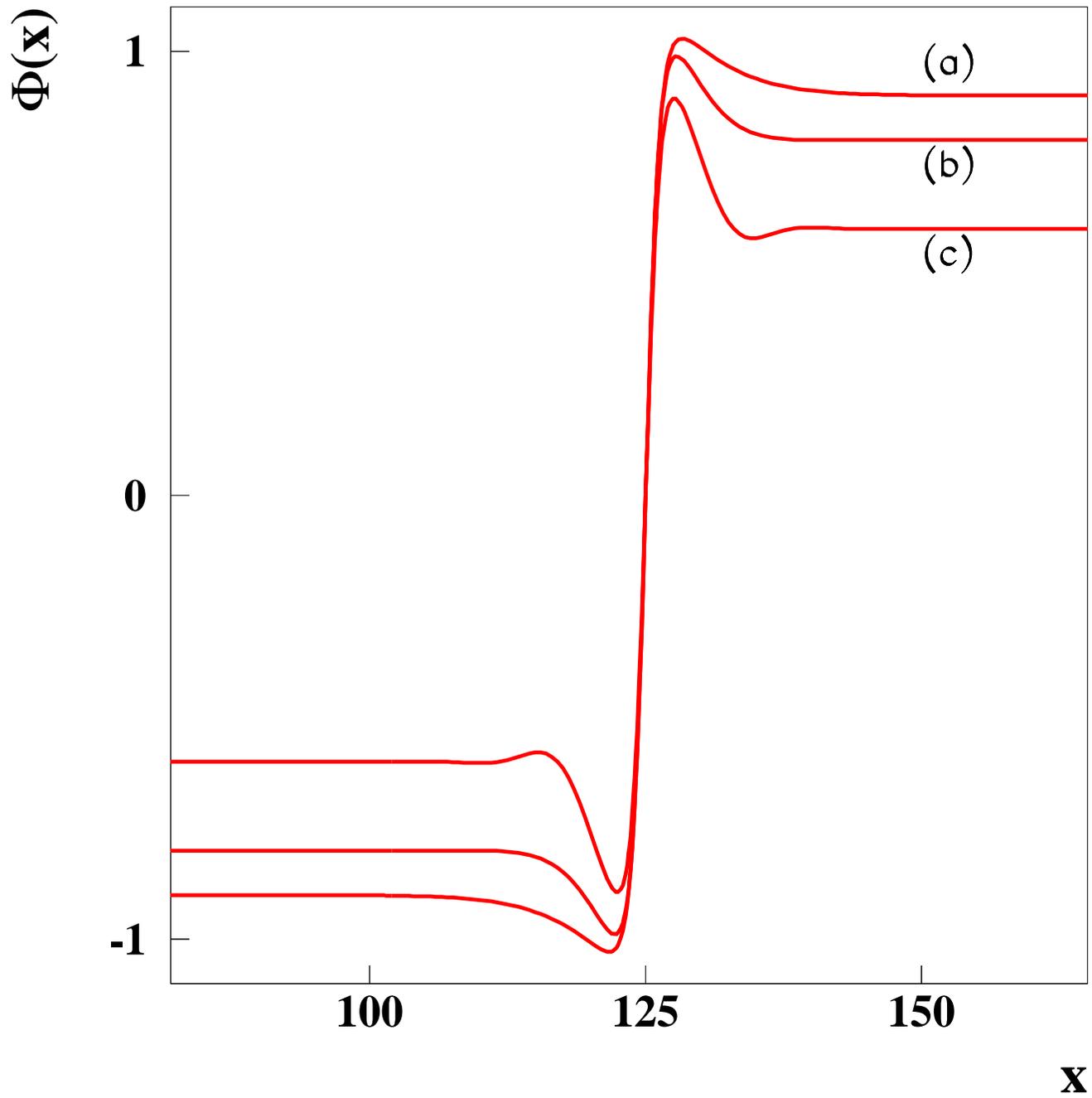
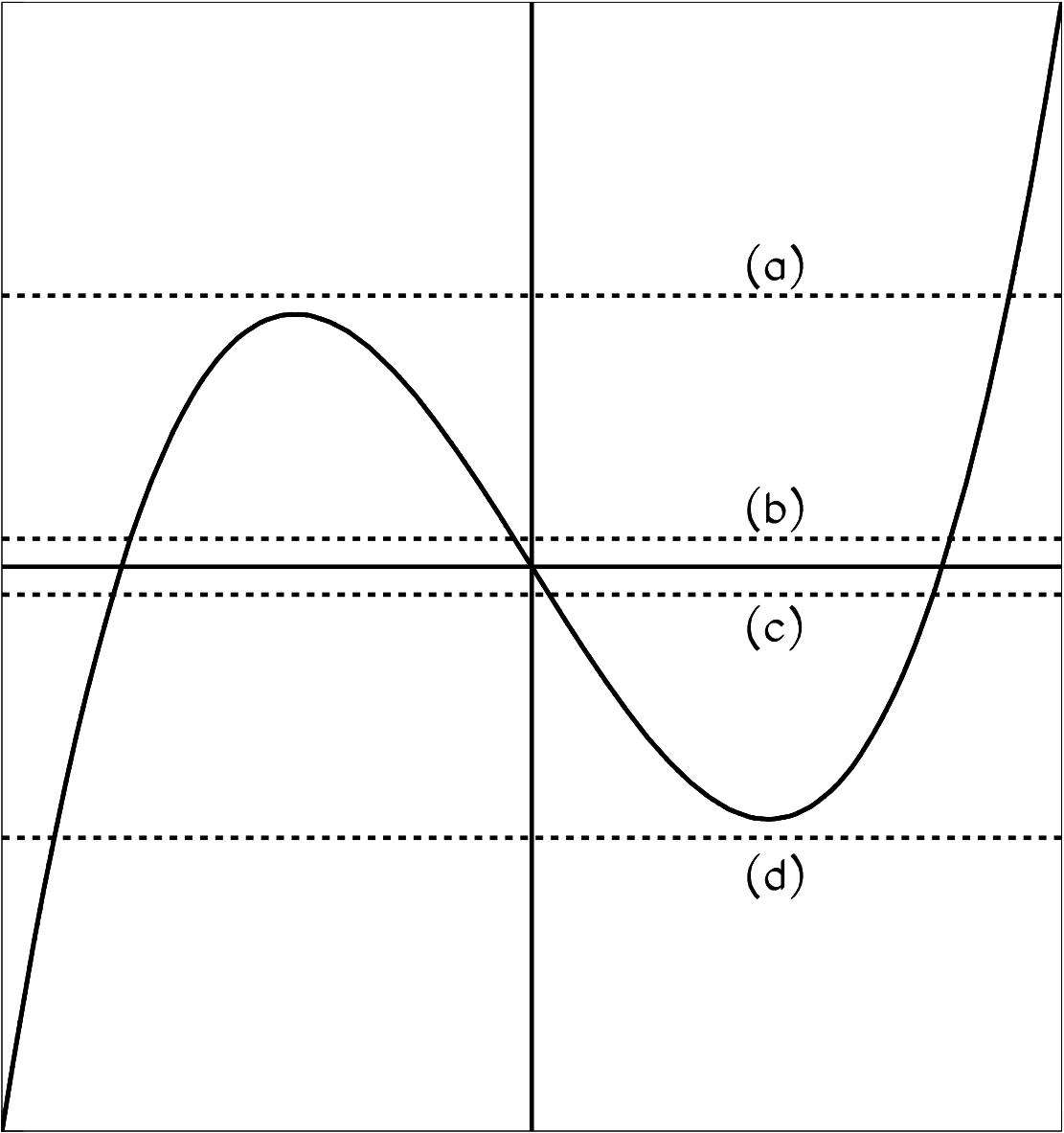
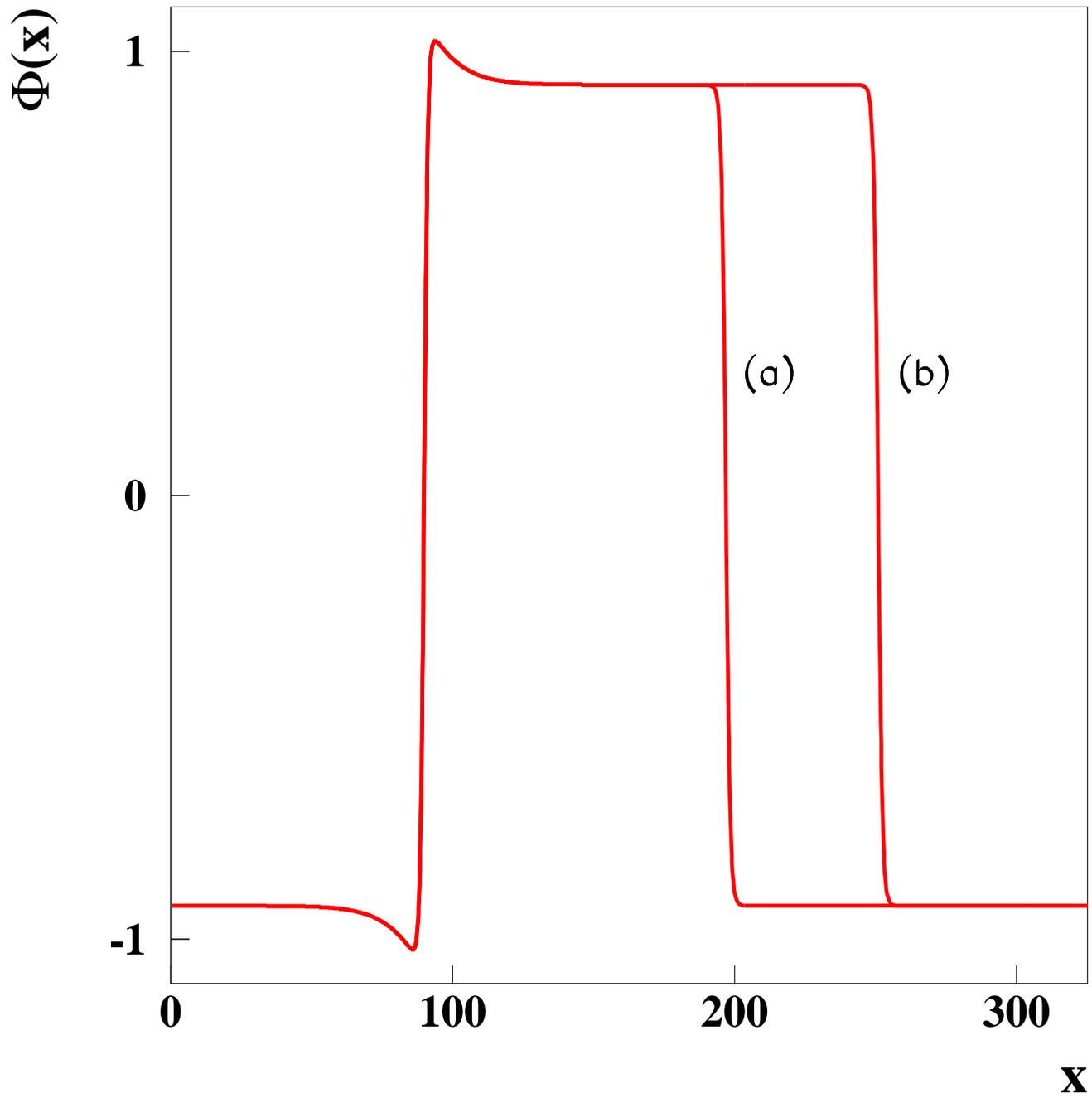
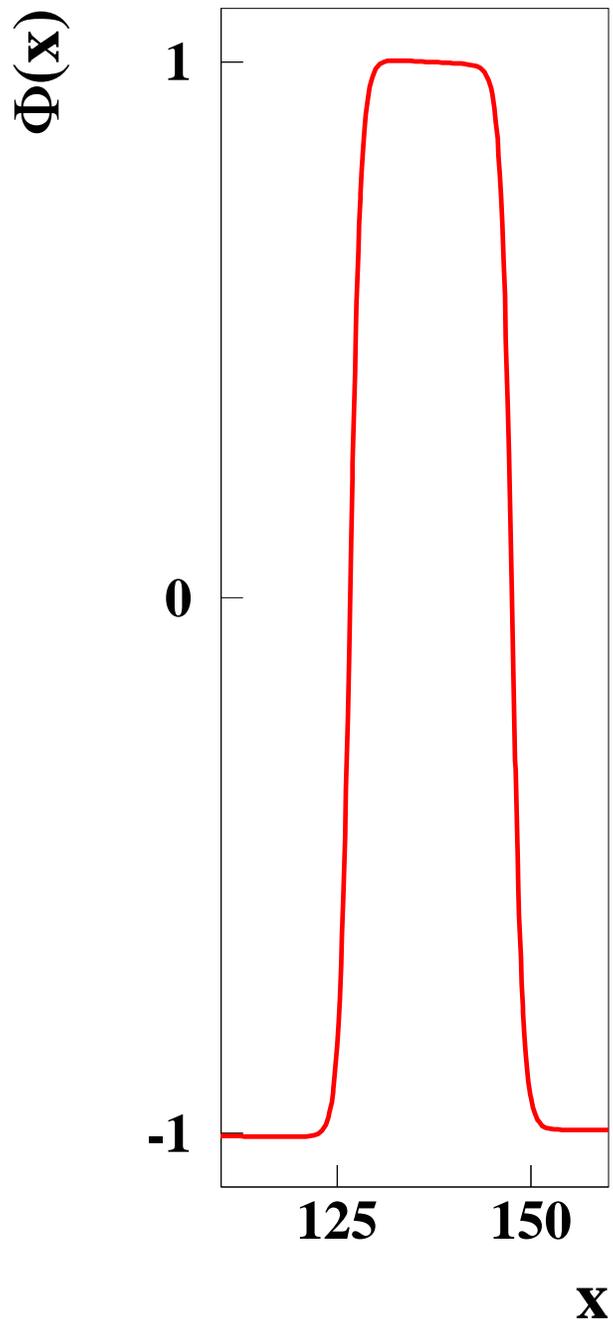
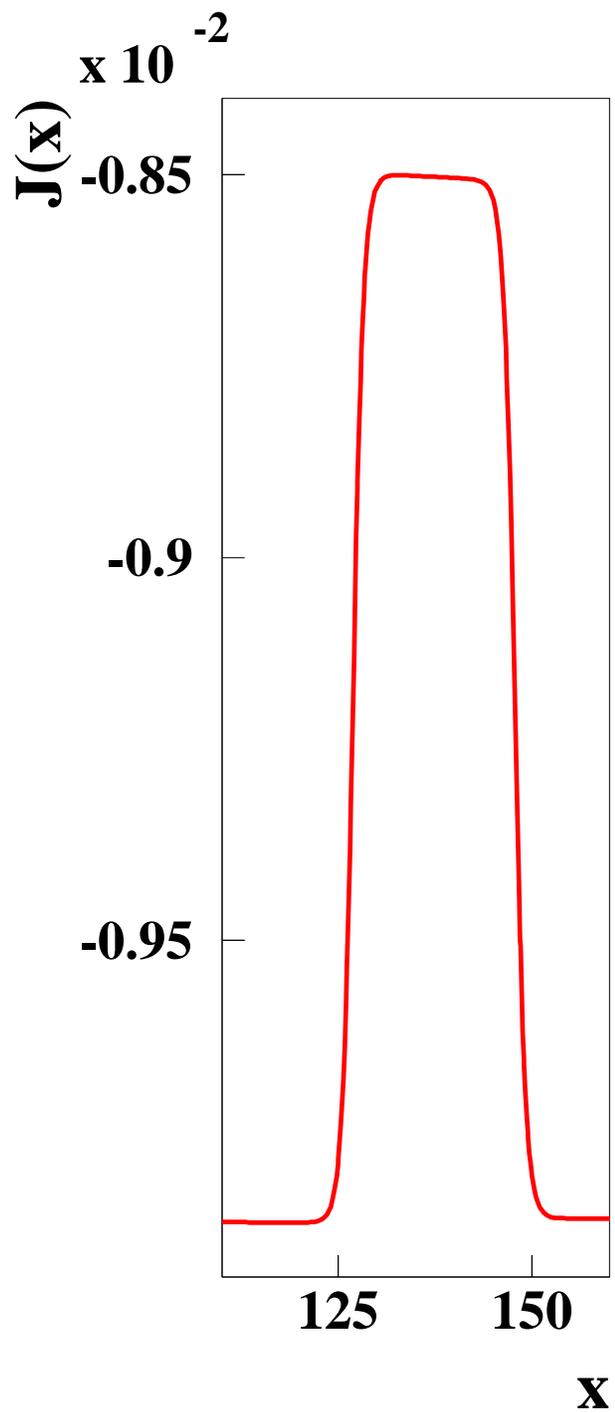
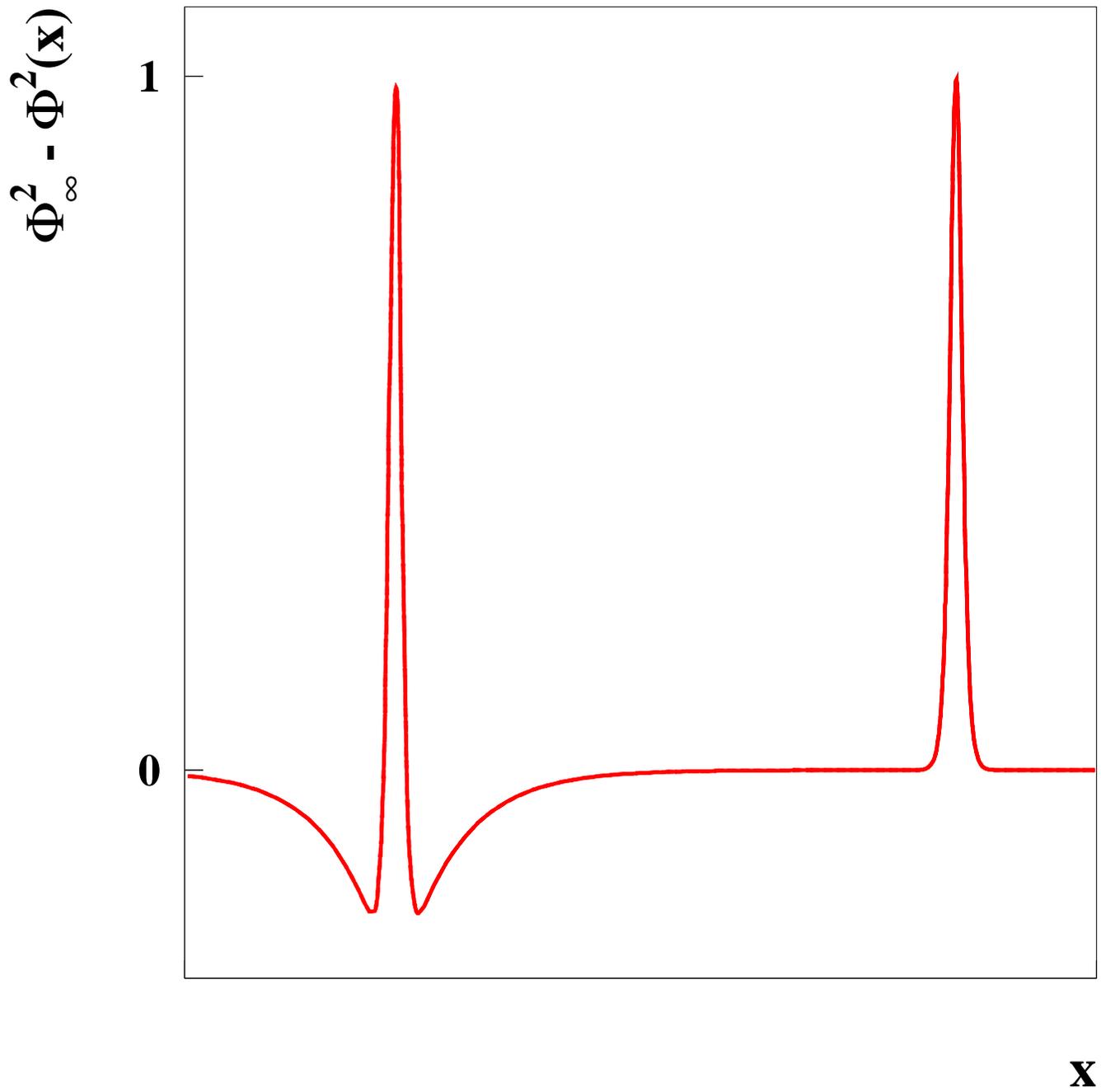
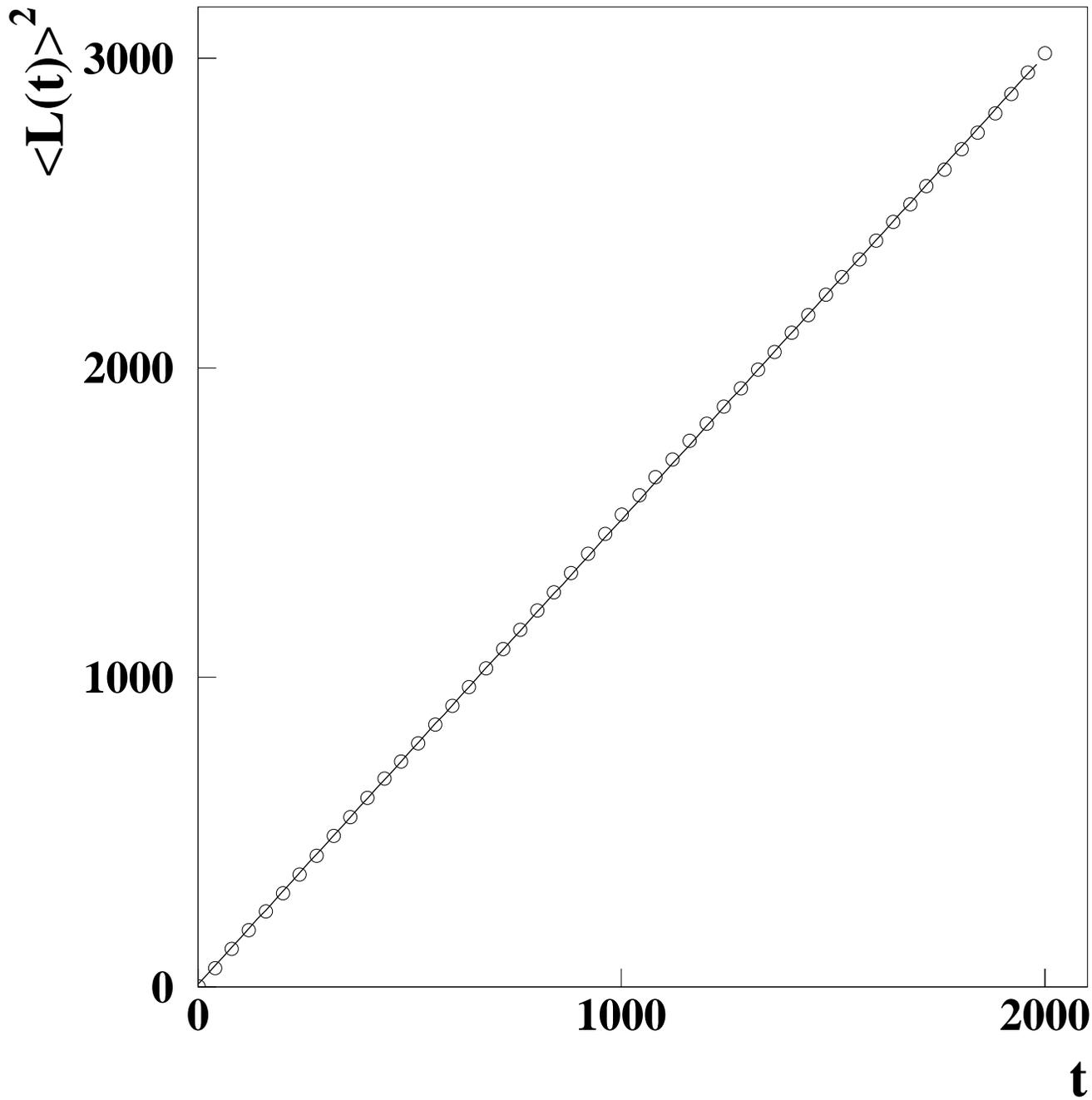
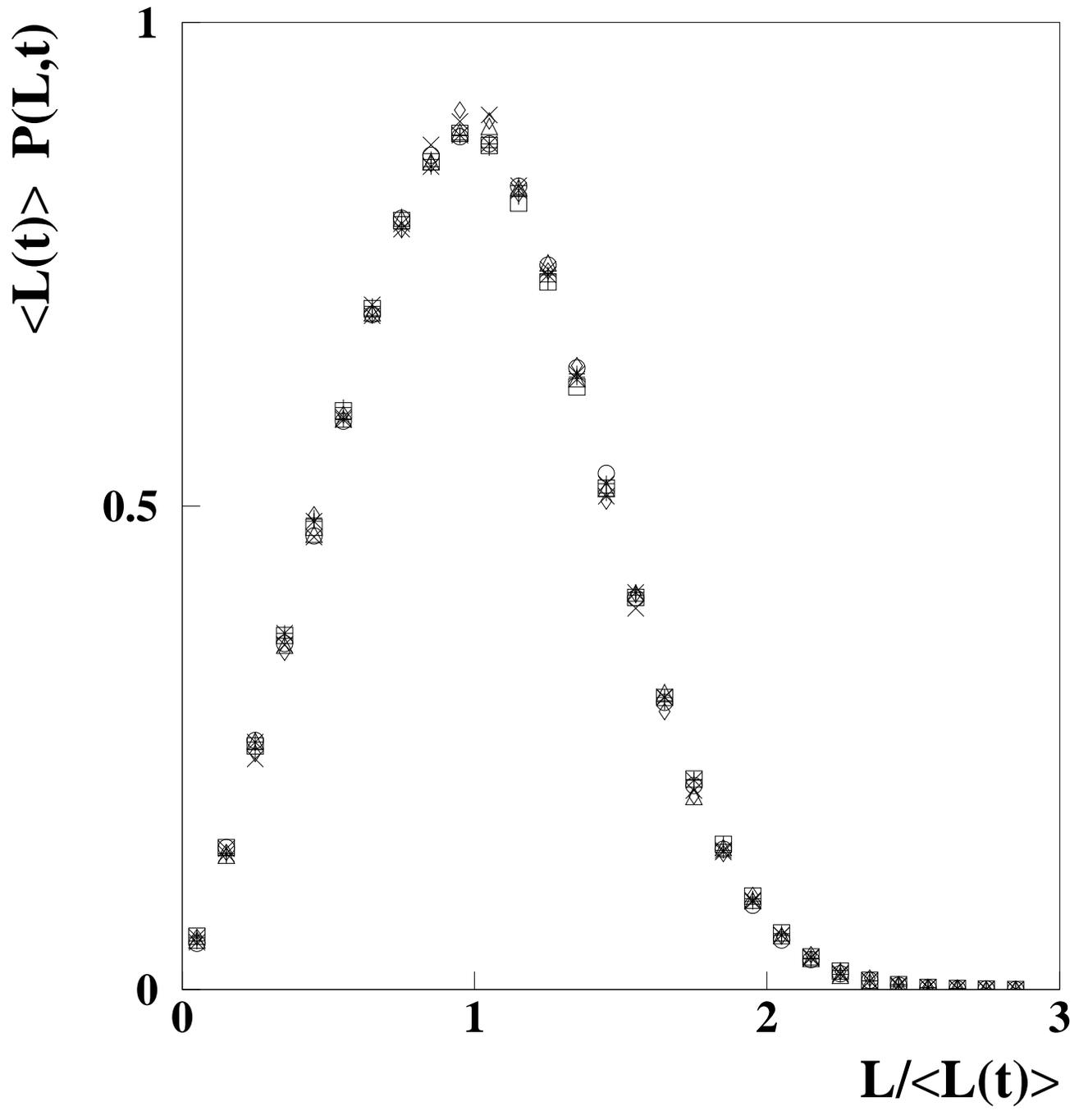
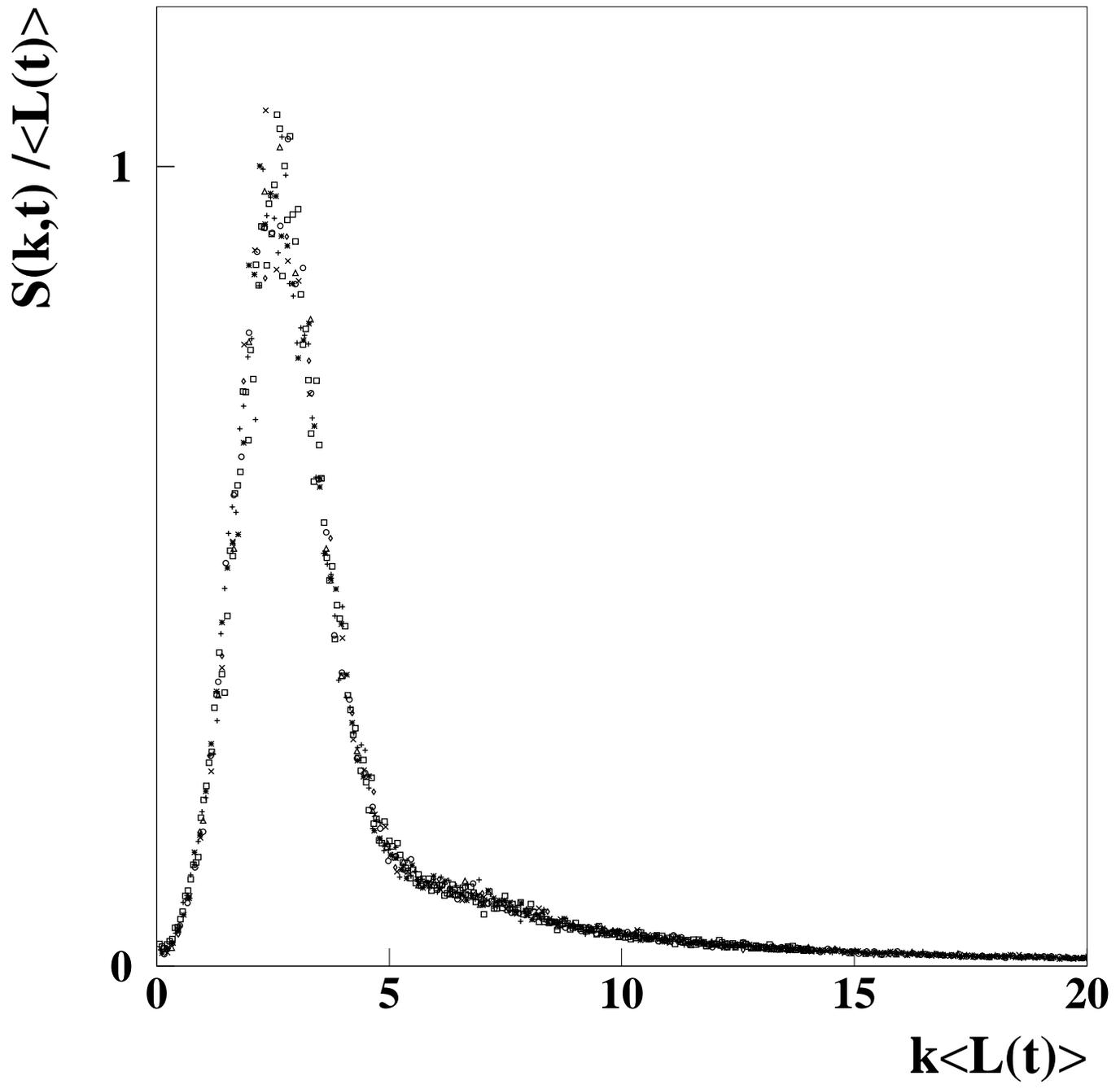
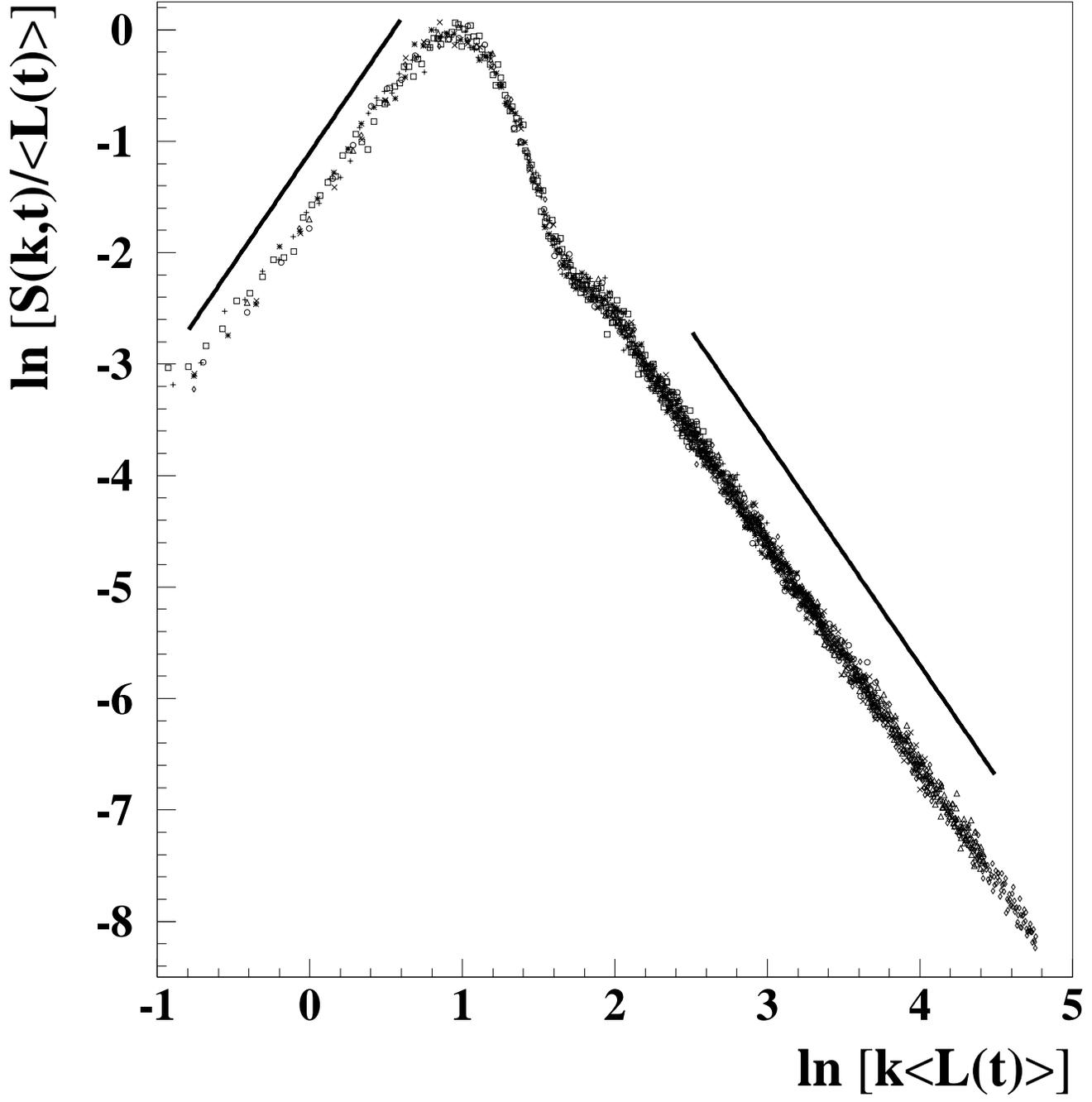
\end{document}